# Multiple Hot-Carrier Collection in Photo-Excited Graphene Moiré Superlattices

Sanfeng Wu[1], Lei Wang[2,3], You Lai[4], Wen-Yu Shan[5], Grant Aivazian[1], Xian Zhang[2], Takashi Taniguchi[6], Kenji Watanabe[6], Di Xiao[5], Cory Dean[7], James Hone[2], Zhiqiang Li[4,*], Xiaodong Xu[1,8,*]

[1]*Department of Physics, University of Washington, Seattle, Washington 98195, USA*
[2]*Department of Mechanical Engineering, Columbia University, New York, NY 10027, USA*
[3]*Department of Electrical Engineering, Columbia University, New York, NY 10027, USA*
[4]*National High Magnetic Field Laboratory, Tallahassee, Florida 32310, USA*
[5]*Department of Physics, Carnegie Mellon University, Pittsburg, Pennsylvania 15213, USA*
[6]*National Institute for Materials Science, 1-1 Namiki, Tsukuba 305-0044, Japan*
[7]*Department of Physics, Columbia University, New York, NY 10027, USA*
[8]*Department of Materials Science and Engineering, University of Washington, Seattle, Washington 98195, USA*

**Abstract:** In conventional light-harvesting devices, the absorption of a single photon only excites one electron, which sets the standard limit of power-conversion efficiency, such as the Shockley-Queisser limit. In principle, generating and harnessing multiple carriers per absorbed photon can improve the efficiency and possibly overcome this limit. Here, we report the observation of multiple hot-carrier collection in graphene/boron-nitride Moiré superlattice structures. A record-high zero-bias photoresponsivity of 0.3 A/W, equivalently, an external quantum efficiency exceeding 50%, is achieved utilizing graphene's photo-Nernst effect, which demonstrates a collection of at least 5 carriers per absorbed photon. We reveal that this effect arises from the enhanced Nernst coefficient through Lifshtiz transition at low-energy Van Hove singularities, which is an emergent phenomenon due to the formation of Moiré mini-bands. Our observation points to a new means for extremely efficient and flexible optoelectronics based on van der Waals heterostructures.

## INTRODUCTION

The Shockley-Queisser efficiency limit for converting light into electricity exists mainly due to the spectral loss of photon energy below and above the absorber's bandgap. For instance, any excess energy of an above-bandgap excitation will dissipate as loss in conventional devices. However, if this excess energy can be utilized to excite secondary carriers, leading to carrier multiplication, it is possible to outperform the standard limit (*1*). The realization of this goal hinges on the identification of suitable physical systems, where both the multiplication process occurs and the subsequent carrier extraction efficiency is high. Among the promising candidates, such as quantum dots or carbon nanotubes, graphene has an outstanding potential because of its broadband absorption, ultrahigh carrier mobility, and structural flexibility. Recent ultrafast optical and photoemission measurements have revealed multiple hot-carrier generation per photon excitation in moderately doped graphene due to strong carrier-carrier scattering (*2–6*). In principle, this process minimizes the spectral loss because graphene has zero-gap and all the absorbed photons release their energy efficiently by generating useful hot carriers, making graphene an excellent platform for light-harvesting. However, graphene optoelectronic devices have not yet shown effective conversion of these hot-carriers into current, with typical zero-bias photoresponsivities ζ of less than 10 mA/W (*7*) implying less than one collected carrier per absorbed photon.

In this article, we demonstrate the collection of multiple hot-carriers upon the absorption of one photon in graphene by creating critical spectral points in the electronic bands via engineered van der Waals heterostructures. In general, Fermi surface topology at band critical points can undergo sudden changes, giving rise to electronic topological transition (*8*), or Lifshitz transition. This effect can cause anomalies in material's properties, such as the conductivity, specific heat, and thermoelectric coefficients (*8*). The latter is particularly important to graphene's optoelectronic behaviors, because its photocurrent generation is dominated by the photo-thermoelectric effects (*9–11*). Away from the Dirac points, the energy spectrum of graphene contains such critical points at which Van Hove singularities (VHSs) appear (*12*). However, the extremely high doping required to reach these VHSs has prevented their experimental access in pristine graphene. Alternatively, the formation of Moiré superlattices in twisted graphene bilayers (*13*) and graphene on hexagonal boron-nitride (h-BN) heterostructures (*14–18*) generates electronic mini-bands that mimic graphene's energy spectrum but with reduced energy scale, providing a unique opportunity to study a variety of physics previously inaccessible (*14–16*, *19*, *20*). In particular, the long-wavelength Moiré superlattice formed in aligned graphene/h-BN heterostructures (Fig. 1A and B) simultaneously achieves high-degree band engineering and excellent transport performance (*14–16*). It has been demonstrated that this superlattice can open up a bandgap in graphene (*16*, *19*), make the two degenerate electronic valleys topologically distinguishable (*20*), generate secondary Dirac points (sDP) (*18*), and give rise to a fractal quantum structure known as Hofstadter's butterfly (*14–16*). Here we reveal the remarkable influence of the emergent low energy VHSs in the superlattice mini-bands on graphene's optoelectronic response, yielding a highly efficient photocurrent generation that may lead to a new type of graphene optoelectronics.

### RESULTS

Our devices were made with graphene encapsulated between h-BN sheets sitting on a graphite gate using recently developed polymer-free transfer techniques(*21*). In each device, large area (~100 μm$^2$) with a clean region in the heterostructure was achieved and fabricated into an edge-contacted Hall-bar geometry(*21*). This enabled spatially resolved photocurrent measurements on a high quality sample with characterization of both longitudinal and Hall resistivity. Figures 1A and C show the schematic diagram and an optical micrograph of a fabricated device. The room temperature mobility of such devices is phonon-limited (*21*) and typically 100,000 cm$^2$/Vs at a carrier density ~ $10^{12}$ cm$^{-2}$. Our measurements were performed at 4.2 K with tunable magnetic field *B* perpendicular to the sample unless otherwise mentioned.

Figure 1D shows the longitudinal resistance $R_{xx}$ of the device, as a function of graphite back-gate voltage $V_g$, measured at *B* = - 50 mT. The on-set peak with resistance ~ 80 KΩ at $V_g$ = - 0.1 V corresponds to the main Dirac point (DP). Two additional resistance peaks are present at

$V_g$ = - 4.7 V (stronger) and + 4.4 V (weaker), representing the secondary Dirac points, the hallmarks of the formation of a long-wavelength Moiré superlattice (*14–17*). Following ref. *15*, we determine the Moiré wavelength to be ~ 14 nm in this device.

The photocurrent measurements were performed under CW laser excitation (660 nm) with about 2 μm beam spot size at the sample. While grounding the source and floating all the voltage probes, we collected photocurrent from the drain contact (Fig. 1A) (See Materials and Methods). Figure 1E plots the observed photocurrent as a function of $V_g$ under weak magnetic fields when placing the laser spot at a selected graphene edge (Fig. 1A). The incident power was set at 1 μW before the objective. We found that the observed photocurrent has a strong dependence on $V_g$. For a slightly nonzero *B* (~ 50 mT), greatly enhanced photocurrent, indicated by the red arrows, is observed nearby both electron- and hole-side sDP (e/h-sDP). We have repeated the observation of this enhancement in a second superlattice device with a Moiré wavelength (~ 10 nm), as shown in Fig. S1. These photocurrents switch direction when *B* is reversed. They appear only when pumping near the device edges and flow oppositely at opposite edges. This chiral edge nature is clearly revealed through a spatially resolved scanning photocurrent map (Fig. 1F) at a selected gate voltage.

Such photocurrents have been attributed to the photo-Nernst effect (*22*). Due to the weak electron-phonon interactions in graphene, optically excited electrons are known to release their energy at a sub-picosecond timescale by generating multiple hot-carriers near the Fermi surface (*2*, *23*, *24*). These processes, leading to photo-thermoelectricity (*9–11*), are largely responsible for the photocurrent generation in various graphene optoelectronic devices (*7*). In our case, because the laser beam is focused onto the sample edge, a temperature gradient of hot carriers forms from edge to bulk. Under the perpendicular magnetic field, a net transverse magneto-thermoelectric current, or Nernst current, can thus flow along the edge (*22*).

What is unique in our observation is that the photo-Nernst current is drastically enhanced near the sDPs (Fig. 1E). To understand this, we carefully inspect the Moiré superlattice electronic band structure (*25*). Although the exact parameters in the Hamiltonian vary from device to device, a general treatment considering the Moiré superlattice with zero twist-angle (*25*) can give an intuitive picture underlying our observation. Figure 2A plots the calculated lowest four bands at one valley and the dashed black arrows indicate sDPs in two selected bands. The corresponding electronic density of states (DOS) is shown in Fig. 2B, where DOS minima feature the DP and sDPs. Both e- and h-sDPs are flanked by two pronounced DOS peaks at each, as labeled by a-d. Those four peaks have been identified as VHSs existing in the Moiré minibands (*25*, *26*).

Our calculation suggests that these VHSs arise from saddle point formation in the Moiré minibands, indicated by the white solid arrows in Fig. 2A. In Fig. 2C we plot the constant energy contour near the saddle point b, where one can clearly see the transition from sDP (μ points, center of blue contour) to DP (center of red contour). The saddle point singularities are located in between μ and κ points that are the local energy minima at the corner of superlattice Brillouin zone (sBZ), as shown by the zoomed-in plot in Fig. 2C. Similar electronic structures can also be found for saddle points a, c and d (Fig. S2).

These band singularities feature Lifshitz transitions, where the electron Fermi surface has a sudden topology change. One direct consequence of such Lifshitz transitions is the orbital switching from electron-like to hole-like (27). This is exactly what happens in the Moiré minibands. Fig. 3a illustrates the evolution of the simulated Fermi surface (See Methods) as the Fermi energy increases from h-sDP to DP, passing through saddle points B (Fig. 2B). Directly above the h-sDP, there are six electron-like Fermi-pockets at μ points. As the Fermi energy increases, the Fermi-surface enlarges but remains electron-like although new electron-like pockets at κ points appear. At the VHS, the separated Fermi pockets connect, after which a single hole-like pocket forms at the sBZ center.

It has been established that Lifshitz transition causes anomalies in a wide range of material's properties, which can be used to identify the appearances of VHSs. In our case, this can be seen in the transverse ($\sigma_{xy}$) and longitudinal ($\sigma_{xx}$) conductivities. Figure 3B shows the $\sigma_{xy}$ at 455 mT as a function of $V_g$. There are sudden sign changes in $\sigma_{xy}$ near the sDPs, showing the carriers experience orbital switching upon slightly tuning the Fermi energy (Fig. 3A). Accompanying these sign changes, pronounced peaks in $\sigma_{xx}$ (Fig. 3C) also appear, particularly the b and c peaks that can be easily identified. These observations reveal the formation of VHSs. Peaks a and d are not observed. We suspect that they appear outside the achievable range of gate voltage. Similar behaviors in conductivities have been theoretically discussed in pristine graphene (28). However, in that case it requires extreme doping to tune the Fermi level up to VHSs, preventing experimental observations.

Intriguingly, it is such a Lifshitz transition leading to the enhanced photo-Nernst current. Figure 3D plots the observed photocurrent $I_{pc}$ as a function of gate voltage. One can see that the conductivity peaks map one-to-one with the enhanced features in $I_{pc}$, unambiguously connecting the photocurrent anomalies to the VHSs. The slight offset in gate voltage at peak b between $I_{pc}$ and $\sigma_{xx}$ may be due to optical gating of graphene through BN (29). We note that, near e-sDP, $I_{pc}$ appears with opposite polarity with respect to the response at VHSs. The nature of this current is not clear, however our simulation of band structures implies that this could result from the presence of two sets of closely located Dirac-like points with broken inversion symmetry (Fig. S3).

To further reveal the underlying physics of $I_{pc}$, we formulate the short-circuit current by $I_{pc} = S_{xy} \langle \Delta T_{el} \rangle / \rho_{xx} \propto S_{xy} / K_{th} \rho_{xx}$ where $S_{xy}$ is the transverse thermoelectric power, $\langle \Delta T_{el} \rangle$ is the average electronic temperature difference from hot center to cold bulk, $\rho_{xx}$ is the longitudinal resistivity and $K_{th}$ is the electron thermal conductivity. If the Wiedemann-Franz law holds, then $K_{th} \propto 1/\rho_{xx}$. As a result, $I_{pc} \propto S_{xy} \equiv N\,B$, where N is the Nernst coefficient and determines the photocurrent response. At the onset of the Lifshitz transitions, the thermoelectric response is significantly enhanced, leading to a large photo-Nernst effect. To reveal this, we compute $S_{xy}$ using the standard Mott formula(*30, 31*), $S_{xy} \propto (\sigma^{-1})_{xx} \left(\frac{\partial \sigma}{\partial E_f}\right)_{xy} + (\sigma^{-1})_{xy} \left(\frac{\partial \sigma}{\partial E_f}\right)_{yy}$, where $E_f$ is Fermi energy. The calculated $S_{xy}$ in Fig. 3E matches $I_{pc}$ well, except for the exact DP and h-sDP, where the Mott formula might be invalid(*31, 32*). This agreement between $S_{xy}$ and $I_{pc}$ confirms the Nernst nature of photocurrent. Moreover, we can conclude that the enhanced $I_{pc}$ is a direct manifestation of the large Nernst coefficient at the VHSs, enabling very efficient extraction of the photo-carriers in graphene Moiré superlattices.

Remarkably, the observed photocurrent corresponds to a giant photoresponsivity $\zeta$. As show in Fig 1E, at *B* = - 50 mT, the maximum photocurrent appears already as large as 200 nA with about 1 μW incident power. This corresponds to $\zeta$ = 0.2 A/W, two orders of magnitude larger than the previously reported photo-Nernst current(*22*), and 20 times larger than the highest reported values in graphene photodetectors at short-circuit and normal-incidence condition(*7*). Figure 4 shows $\zeta$ at $V_g$ = 4 V as a function of *B* field (see also Fig. S4). One can see that $\zeta$ increases linearly below 0.1 T and then reaches its maximum value of about 0.3 A/W at 0.25 T before it decreases as a function of fields. We note that in the classical region, the Nernst current is proportional to *N·B*. The linear increase of photocurrent with *B* is hence consistent with Nernst mechanism while the decrease with *B* at high field implies the transition of the system into quantum regime (Fig. S4).

Another figure of merit of an optoelectronic device is its external quantum efficiency η, the ratio between the number of collected carriers and that of the incident photons. It can be formulated by η = $\frac{I_{pc}}{e}$ / $\frac{p}{\hbar \omega}$ = $\zeta \hbar \omega$/e, where $I_{pc}$ is photocurrent amplitude, p is excitation power, $\hbar \omega$ is the photon energy, and e is the electron charge. It yields a maximum η over 50% in this device. Consequently, the internal quantum efficiency in this device greatly exceeds unity since graphene's absorption coefficient α is much less than one. Although the superlattice structure may modulate its exact value, α is expected to be on the order of 2.3%, with an upper limit of about 10% (*33*). This implies that upon the absorption of one photon, at least η/α > 5 electrons are collected. Considering the laser only illuminates the graphene edge, the effective excitation power on the sample is smaller than measured. We expect that the actual captured hot-carriers per photon absorption are much larger than the number estimated above.

**DISCUSSION**

In summary, our results point to a unique way to harness VHSs emergent in Moiré band engineering for converting photo-excited hot-carriers into current in van der Waals heterostructures. The Lifshitz transitions at mini-bands VHSs facilitate enhanced thermoelectricity when Fermi level is aligned with the singularity, leading to the multiple hot-carrier collection and the subsequent giant photocurrent generation. Although the present study employs the visible light excitation, we also expect similar effect at other wavelengths, such as in the near and mid-infrared region, given that the photo-excitation is above the secondary Dirac point.

In the current device geometry, we have exploited photo-Nernst effect, where a small magnetic field is required for generating current. On one hand, fabrication of the graphene/h-BN superlattice on a ferromagnetic layer may overcome this limitation(*34*), making possible optoelectronic devices based on photo-Nernst effects. On the other hand, photo-Nernst geometry is in principle not necessary. Future work may develop devices based on PN junctions built on the Moiré superlattice structure, where the effect of VHSs on longitudinal thermoelectric coefficient is also expected. Another limitation in the present study may be the need of low temperature. The reduced carrier mobility and broadened spectrum at VHSs at high temperature might be potential limitations for the device performance. We anticipate that further device engineering to optimize graphene's electronic and thermal properties will be necessary for possible multiple hot carrier collection at room temperature.

### MATERIALS AND METHODS

Photocurrent measurements

In the measurements, a 660 nm CW laser is focused on the sample with about 2 μm spot size by an optical microscope objective (90% transmission). Both sample and objective are loaded together into a continuous helium flow superconducting magnet (17.5T). The temperature is held at 4.2 K. The sample is mounted on an attocube nano-positioning stage, which can determine the laser exposure region. The intensity of the laser beam is modulated at 800 Hz by a mechanical chopper. The photocurrent is measured by a lock-in preamplifier at a reference frequency (~800 Hz).

Resistance measurements

We measured both longitudinal $R_{xx}$ and transverse $R_{xy}$ through standard Hall-bar techniques. We apply 1 mV excitation with 10 Hz oscillating frequency to source and measure the drain current $I_d$ by a lock-in preamplifier. The voltage drops between the two longitudinal probes $V_{xx}$ and between the two transverse probes $V_{xy}$ are recorded respectively. Resistances were obtained using $R_{xx} = V_{xx}/I_d$ and $R_{xy} = V_{xy}/I_d$. Since the channel width and length are equal in our devices, the conductivities were obtained using $\sigma_{xx} = R_{xx}/(R^2_{xx} + R^2_{xy})$ and $\sigma_{xy} = -R_{xy}/(R^2_{xx} + R^2_{xy})$.

Band structure simulations

The miniband structure of graphene on a hexagonal substrate is calculated using the following Hamiltonian(*25*)

$$H = v\boldsymbol{p}\cdot\boldsymbol{\sigma} + u_0\varepsilon_0 f_1(\boldsymbol{r}) + u_1\left(\frac{\varepsilon_0}{b}\right)[\boldsymbol{l}_z\times\boldsymbol{\nabla} f_2(\boldsymbol{r})]\cdot\boldsymbol{\sigma}\tau_z + \varepsilon_0 u_3 f_2(\boldsymbol{r})\sigma_z\tau_z + \varepsilon_0 \tilde{u}_3 f_1(\boldsymbol{r})\sigma_z\tau_z$$

where **σ** and **τ** operate in the sublattice and valley pesudospin space, respectively. The first term describes the original Dirac points in pristine graphene. The next three terms with dimensionless parameters $u_0$, $u_1$, and $u_3$ describe the overall scalar potential modulation, modulated nearest-neighbor hopping, and inversion-symmetric sublattice potential, respectively. The structural factor are given by $f_1(r) = \sum_{m=0}^{5} e^{i\mathbf{b}_m \cdot \mathbf{r}}$ and $f_2(r) = i\sum_{m=0}^{5}(-1)^m e^{i\mathbf{b}_m \cdot \mathbf{r}}$, where $\mathbf{b}_m$ are the reciprocal lattice vectors of the Moiré superlattice. The last term, proportional to $\tilde{u}_3$, breaks the inversion symmetry between the A and B sublattice. We found that a small inversion-symmetry-breaking perturbation $\tilde{u}_3$ is important to capture the features around the e-sDP, particularly the small bump in DOS between the VHS peak c and d. A characteristic energy $\varepsilon_0 = \hbar v b$ is introduced, where $b = |\mathbf{b}_0| \sim (4\pi/3a)\sqrt{(\delta^2 + \theta^2)}$ with 1+ $\delta$ being the ratio of lattice constants with and without substrate and $\theta$ the twisting angle. Throughout our calculation $\theta$ has been set to zero. Diagonalizing the Hamiltonian yields the miniband structure and the DOS. We find that the main feature of the experimental result can be qualitatively explained with only two nonzero parameters, $u_0$ and $\tilde{u}_3$. The actual calculation is performed with $(u_0, u_1, u_3, \tilde{u}_3) = (-0.15, 0, 0, 0.004)$.

## SUPPLEMENTARY MATERIALS

Fig. S1. Data taken from another graphene/BN superlattice device.

Fig. S2. Saddle points corresponding to VHS peaks a, c and d.

Fig. S3. Band structure at e-sDP.

Fig. S4. Raw data corresponding to Fig. 4 in main text.

**Acknowledgments:**

This work was mainly supported by the National Science Foundation (NSF, DMR-1150719) and AFOSR (FA9550-14-1-0277). The measurements were performed at the National High Magnetic Field Laboratory (NHMFL), which is supported by National Science Foundation Cooperative Agreement No. DMR-1157490 and the State of Florida. Y.L. is supported by UCGP program at NHMFL. J.H. and L.W. are supported by ONR (N00014-13-1-0662) and INDEX (NERC 2013-NE-2399). X.X. acknowledges the support by Boeing Distinguished Professorship and from the State of Washington–funded Clean Energy Institute.

**Author Contributions**:

X.X., Z.L. and S.W. conceived the experiment. G.A., S.W., Y.L., Z.L. and X.X. designed and built the experimental setup. L.W., X.Z., C.D., and J.H. designed and fabricated the devices. S.W. performed the measurements and analyzed the data. Y.L. and Z.L. assisted on the measurements. W.S. and D.X. provided the theoretical support with the assistance of S.W. and X.X. T.T. and K.W. provided the BN crystals. S.W. and X.X. wrote the paper with the assistance of W.S., D.X., J.H. and C.D. All authors discussed the results and commented on the manuscript.

**Competing Interests:**

The authors declare that they have no competing interests.

**Corresponding authors:**
X.X. (xuxd@uw.edu) or Z.L. (zli@magnet.fsu.edu)

# Figure 1

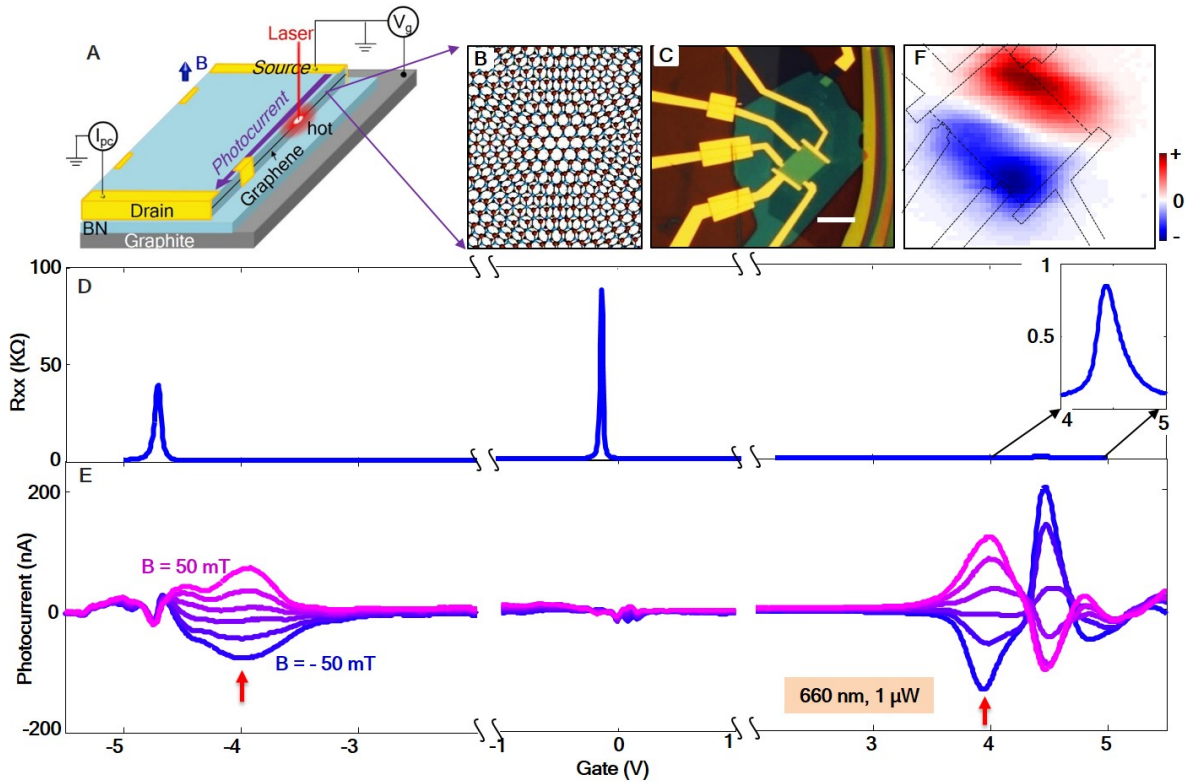

**Fig. 1. Anomalous photo-Nernst effect in graphene/BN superlattices.** (**A**) Schematics of device and photocurrent measurement. Edge-contacted graphene is encapsulated in between *h*-BN sheets sitting on top of a graphite back gate. (**B**) Cartoon depiction of Moiré superlattice when aligning graphene crystal with *h*-BN substrate. (**C**) Optical image of one device. Scale bar: 10 μm. (**D**) Longitudinal resistance $R_{xx}$ as a function of gate at 50 mT showing one DP and two sDPs. Inset zooms in the electron-side sDP peak. (**E**) Photocurrent generation as a function of gate under magnetic field varying from -50 to 50 mT with step size of 20 mT. The drain current is recorded while grounding the source, as shown in (**A**). Laser power is set to be 1 μW before microscope objective. The red arrow indicates the enhanced photocurrent features. T = 4.2 K. (**F**) Typical spatially resolved scanning photocurrent map (taken at 20 K), showing the chiral edge pattern consistent with the photo-Nernst current, which is generated at the two edges with opposite signs.

# Figure 2

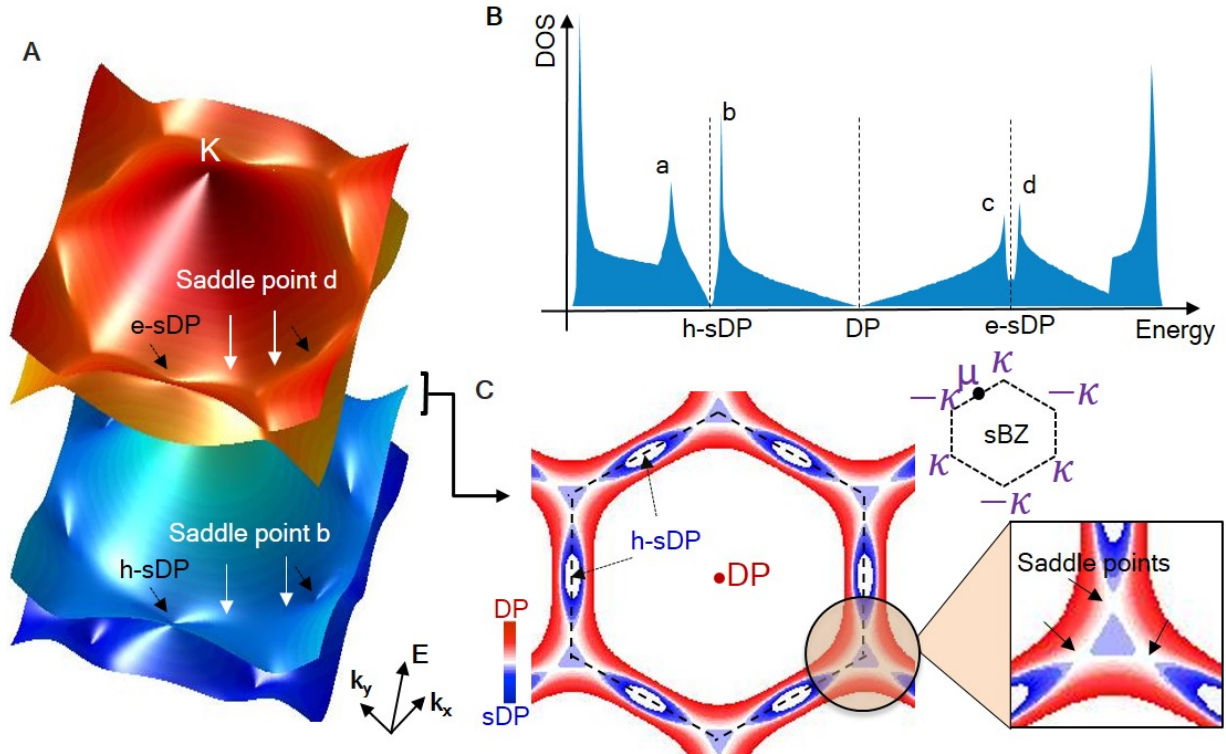

**Fig. 2. Van Hove singularities in Moiré minibands.** (**A**) Simulated lowest four bands in K valley for graphene/BN superlattice with zero twist angle. e/h-sDPs are indicated by the dashed black arrows. The solid white arrows locate the saddle points in the first hole and second electron bands. (**B**) The electronic density of states (DOS) corresponding to the simulated energy bands, showing the saddle point VHSs labeled by a-d in (**A**). (**C**) Constant energy contour of the first hole band in momentum space near the saddle point singularities. The dashed hexagonal lines indicate the sBZ, which is further shown at the top right inset. Red (blue) color denotes the Fermi-surface approaching DP (sDP). In addition to the sDP at μ points, another local energy minima is located at symmetry points κ. The saddle points are located in between μ and κ points, as depicted by zoomed-in plot at bottom right. The formation of saddle point VHSs in superlattice minibands appears in all four bands in (**A**).

Figure 3

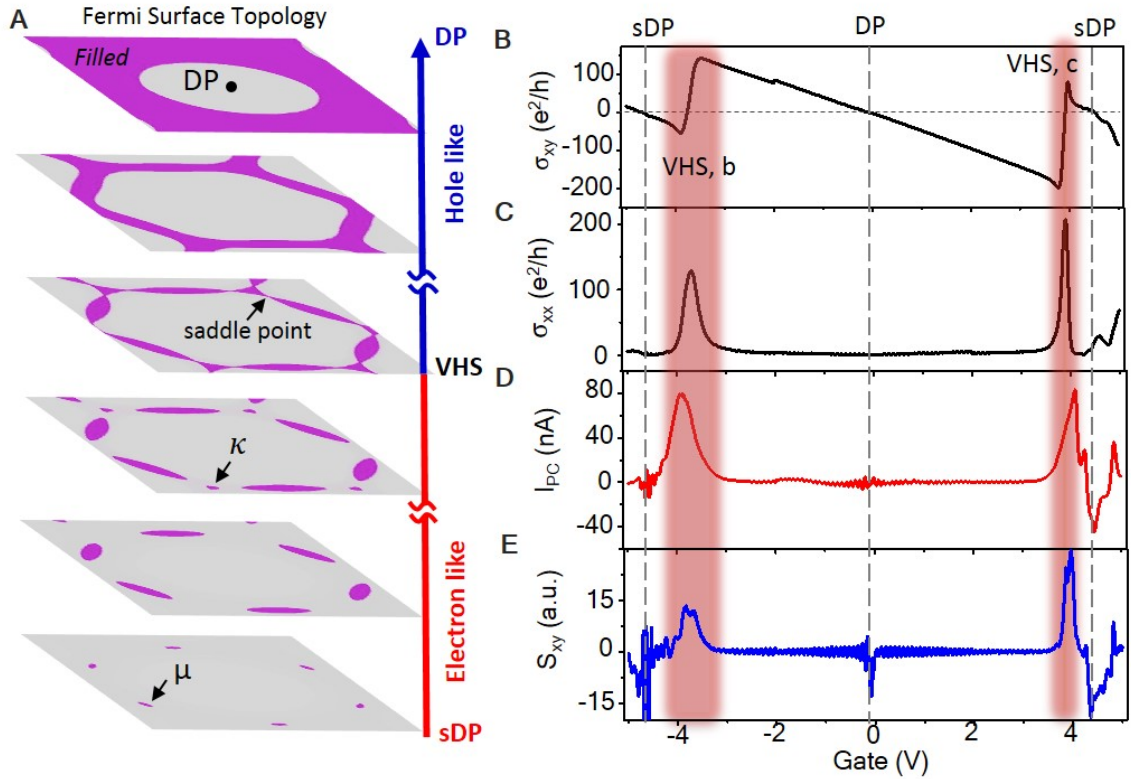

**Fig. 3. Anomalies induced by Lifshitz transitions.** (**A**) Schematic evolution of Fermi surface topology when tuning the Fermi energy from h-sDP to DP, demonstrating the Lifshitz transition. Magenta (grey) color corresponds to filled (empty) region in momentum space. One can see that the orbital of carriers changes from electron-like to hole-like when passing through the saddle point VHSs. Gate dependent (**B**) transverse conductivity $\sigma_{xy}$, (**C**) longitudinal conductivity $\sigma_{xx}$, (**D**) photo-Nernst current $I_{pc}$ with 0.5 µW excitation, and (**E**) transverse thermoelectric power $S_{xy}$ calculated from Mott formula at B = 455 mT. The sudden sign-changes in $\sigma_{xy}$ and the accompanying peaks in $\sigma_{xx}$ demonstrate the formations of the superlattice VHSs. They are unambiguously linked to the anomalous features in $S_{xy}$ and $I_{pc}$. The regions of VHSs in (**B**) and (**C**) are highlighted and the locations of DP and sDPs are also indicated by the gray dashed lines.

# Figure 4

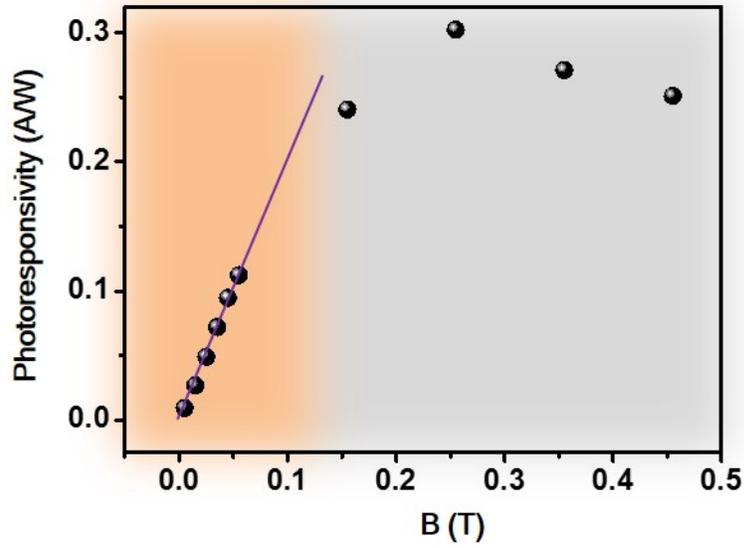

**Fig. 4. Photoresponsivity ζ as a function of B**. ζ linearly increases with B below ~ 0.1T and then slightly decreases at higher fields, indicating a transition from the classical to quantum regime. The solid line is a guide to the eye to the linear response. The data is taken at $V_g = 4$ V.

Supplementary Materials for

**Multiple Hot-Carrier Collection in Photo-Excited Graphene Moiré Superlattices**

Sanfeng Wu[1], Lei Wang[2,3], You Lai[4], Wen-Yu Shan[5], Grant Aivazian[1], Xian Zhang[2], Takashi Taniguchi[6], Kenji Watanabe[6], Di Xiao[5], Cory Dean[7], James Hone[2], Zhiqiang Li[4,*], Xiaodong Xu[1,8,*]

Correspondence to:  to X.X. (xuxd@uw.edu) or Z.L. (zli@magnet.fsu.edu)

**This PDF file includes:**

Fig. S1. Data taken from another graphene/BN superlattice device.

Fig. S2. Saddle points corresponding to VHS peaks a, c and d.

Fig. S3. Band structure at e-sDP.

Fig. S4. Raw data corresponding to Fig. 4 in main text.

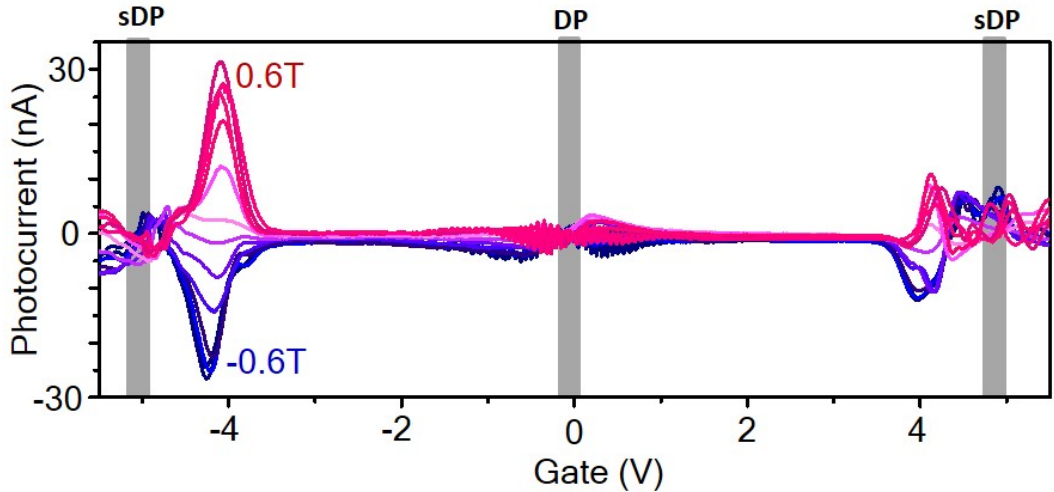

**Fig. S1. Data taken from another graphene/BN superlattice device**. Photocurrent is plotted as a function of gate under selected magnetic field from -0.6 T to 0.6 T, measured by laser exciting graphene edge with 500 nW incident power. Anomalous photo-Nernst current is observed at the vicinity of the sDPs, similar to Fig. 1A in the main text. The sDPs and DP are marked. The corresponding maximum photoresponsivity in this device is 60 mA/W, demonstrating the collection of more than one carrier per photon excitation. The maximum photoresponsivity of this device is smaller than the one presented in the main text, which may be due to the different Moiré wavelengths between the two devices.

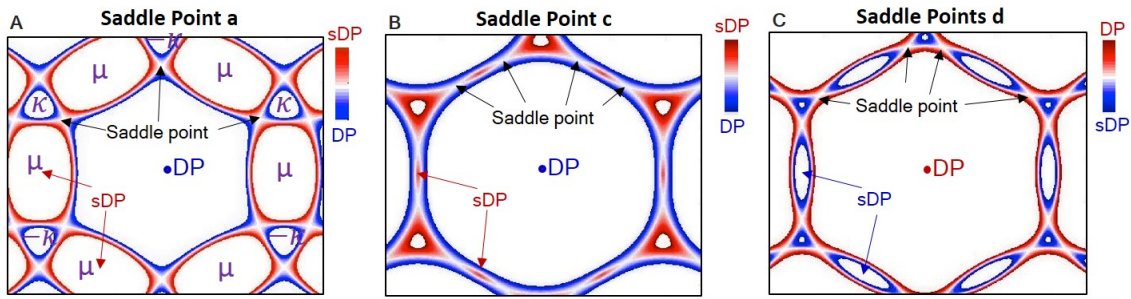

**Fig. S2. Saddle points corresponding to VHS peaks a, c and d.** Calculated constant energy contours in momentum space at the vicinity of the saddle points in the second hole-band (**A**), the first electron-band (**B**) and the second electron-band (**C**), complementary to the Fig. 3 in main text where the second hole-band is shown. The sDPs, DP, symmetry points in sBZ and saddle points are all marked.

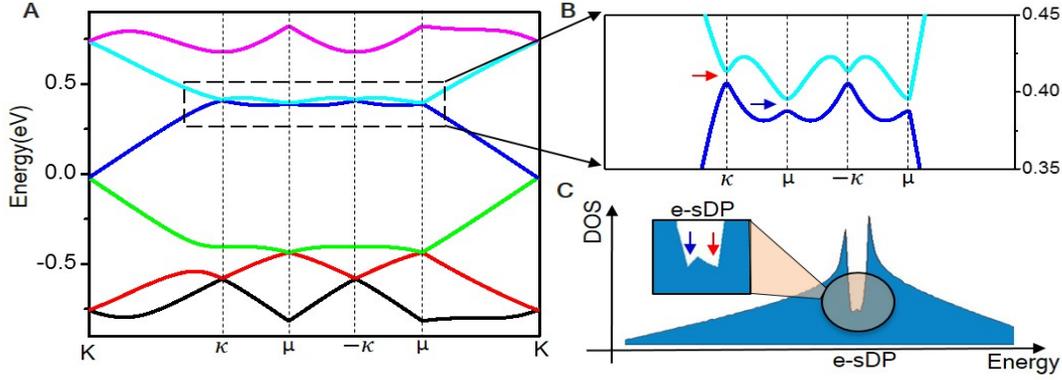

**Fig. S3. Band structure at e-sDP. A**, Calculated band structure for zero-twisted superlattice, using parameters $(u_0, u_1, u_3, \tilde{u}_3) = (-0.15, 0, 0, 0.004)$, where $\tilde{u}_3$ is introduced to take into account broken inversion symmetry. **B** Zoomed-in band structure at e-sDP. One can see that two sets of Dirac-like points appears closely in energy, at $\mu$ (blue arrow) and $\kappa$ (red arrow), respectively. Both open a bandgap due to the broken inversion symmetry. **C**, the corresponding DOS at e-sDP. It shows (1) a non-vanishing DOS at e-sDP and (2) a little DOS bump in between the two VHS peaks. The non-vanishing DOS agrees with the suppressed resistance measured at e-sDP. The little bump agrees with the fine feature at e-sDP appeared in longitudinal conductivity $\sigma_{xx}$ (Fig. 3B), where a small conductivity bump is seen. We suspect that this bump is inherently correlated to the noticeable large photocurrent at e-sDP. At this bump, the carrier orbital switch from hole-like to electron-like, as implied by the $\sigma_{xy}$ plot in Fig. 3C. Consequently, the $S_{xy}$ displays an appreciable negative value (Fig. 3C), so does the observed photocurrent (Fig. 3D), consistent with the general expectations that thermal transport coefficient is sensitive to Fermi surface properties.

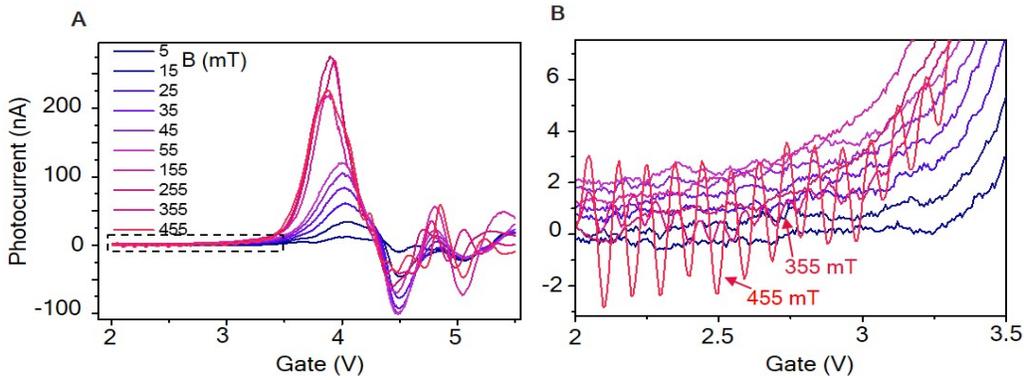

**Fig. S4. Raw data corresponding to Fig. 4 in main text. A,** Magnetic-field dependent photocurrent generation at the vicinity of e-sDP, with 1 μW incident laser power. **B**, Zoomed-in plot of the region indicated by the dashed rectangular in A. It shows the pronounced quantum oscillations at 355 mT and 455 mT, implying the transition to quantum regime at these fields.